\documentclass[fleqn,usenatbib]{mnras}

\usepackage{newtxtext,newtxmath}

\usepackage[T1]{fontenc}
\usepackage{ae,aecompl}

\usepackage{graphicx}    
\usepackage{amsmath}    
\usepackage{amssymb}    
\usepackage[colorinlistoftodos]{todonotes}
\usepackage{color,soul}
\usepackage{float}

\title[\textit{K2}:BS]{\textit{K2}: Background Survey -- the search for undiscovered transients in \textit{Kepler}/\textit{K2} data}

\author[R. Ridden-Harper et al.]{R. Ridden-Harper,$^{1,2,3}$\thanks{E-mail: ryan.ridden-harper@anu.edu.au}
B. E. Tucker,$^{1,2,4}$ M. Gully-Santiago,$^5$ G. Barentsen,$^5$ \newauthor 
A. Rest,$^{3,6}$ P. Garnavich,$^7$ E. Shaya$^8$ 
\\
$^{1}$Research School of Astronomy \& Astrophysics, Mount Stromlo Observatory, The Australian National University,\\ Cotter Road, Weston Creek, ACT 2611, Australia\\
$^{2}$The ARC Centre for All-Sky Astronomy in 3-Dimensions (ASTRO3D)\\
$^{3}$Space Telescope Science Institute, 3700 San Martin Drive, Baltimore, MD 21218, USA.\\
$^{4}$The National Centre for the Public Awareness of Science, The Australian National University,\\ Canberra, ACT 2601, Australia\\
$^5$Bay Area Environmental Research Institute, P.O. Box 25, Moffett Field, CA 94035, USA. \\
$^6$The Johns Hopkins University, Baltimore, MD 21218, USA.\\
$^7$Physics Department, 225 Nieuwland Hall, University of Notre Dame, Notre Dame, IN 46556, USA.\\
$^8$Astronomy Department, University of Maryland, College Park, MD 20742-2421, USA.
}

\date{Accepted XXX. Received YYY; in original form ZZZ}

\pubyear{2019}

\begin{document}
\label{firstpage}
\pagerange{\pageref{firstpage}--\pageref{lastpage}}
\maketitle

\begin{abstract}
The \textit{K2} mission of the \textit{Kepler} Space Telescope offers a unique possibility to examine sources of both Galactic and Extra-galactic origin with high cadence photometry. Alongside the multitude of supernovae and quasars detected within targeted galaxies, it is likely that \textit{Kepler} has serendipitously observed many transients throughout \textit{K2}. Such events will likely have occurred in background pixels, coincidentally surrounding science targets. Analysing the background pixels presents the possibility to conduct a high cadence survey with areas of a few square degrees per campaign. We demonstrate the capacity to independently recover key \textit{K2} transients such as KSN~2015K and SN~2018oh. With this survey, we expect to detect numerous transients and determine the first comprehensive rates for transients with lifetimes $\leq1$~day.

\end{abstract}




\section{Introduction}

Launched in 2009, the \textit{Kepler Space Telescope} (\textit{Kepler}) allowed for high precision and high cadence stellar photometry \citep{Basri2005}, with the goal of detecting thousands of exoplanets \citep{Batalha2014}, until the failure of two reaction wheels led to a new mission profile -- \textit{K2}. The \textit{K2} mission extended \textit{Kepler} observations along the ecliptic, presenting the opportunity to apply high precision and cadence photometry to a multitude of galactic and extragalactic sources \citep{Howell2014}. 

The reduced telescope stability present in \textit{K2} affords it a lower photometric precision than the original \textit{Kepler} mission. Despite the drop in photometric precision, \textit{K2} campaigns C01 to C19 were successful in obtaining unparalleled photometric data that lead to among other things, the detection of new exoplanets \cite[e.g.,][]{Pearson2018}, numerous supernovae (SN) \citep[e.g.,][]{Garnavich2016,Olling2015,Rest2018,Dimitriadis2018,Shappee2018,Li2018}, and variability in quasars \citep{Aranzana2018}.  

The high cadence \textit{K2} observations make it possible to study short-duration structures in supernovae light curves and detect short-duration transients. Analysis of \textit{K2} data has shown potential evidence for SN Ia shock interaction, as described in \citet{Kasen2010}, in at least one SN Ia, SN 2018oh \citep{Dimitriadis2018}, however, not in some other SN Ia \citep{Olling2015}. \textit{K2} has also shown evidence of shock breakouts in core-collapse supernovae (CCSN) \citep{Garnavich2016}, and lead to the discovery of KSN~2015K, a rapid transient that is thought to be powered by stellar ejecta producing a shock in the circumstellar medium \citep{Rest2018}. Short cadence observations were crucial to these results and offer the possibility to analyse a time domain that has previously been technically infesable.

All known \textit{K2} transients were detected through targeted observations of pre-selected galaxy targets, however, this may not be all the transient events detected with \textit{Kepler}. As seen in \citet{Barclay2012} and \citet{Brown2015b} it is possible that \textit{Kepler} has serendipitously detected numerous events in background or sky pixels, such as super-outbursts of dwarf novae. Such background events will likely be faint and will require an analysis that will be heavily influenced by detector noise and telescope stability.

The high cadence observations of \textit{Kepler}, present a unique possibility to probe a new parameter space unavailable to previous transient surveys. Current transient surveys cover larger areas than \textit{Kepler}, however, they have longer cadences. The Pan-STARRS Medium Deep Survey had an nominal cadence of 3 days in any given filter to a depth of $\sim23.3$~mag \citep{Rest2014,Chambers2016}; ZTF surveys the the sky North of $\delta = \rm -31^o$ every 3 nights to $\sim 20.4$~mag, through the public survey \citep{Bellm2019}; ASAS-SN covers the entire sky to $\sim 17$~mag, at a nominal cadence of a few days \citep{Kochanek2017}; and ATLAS covers the sky North of $\delta=-30$ every two nights to $\sim 20.2$~mag \citep{Tonry2018}. Previous surveys, such as the SNLS, ESSENCE, and SDSS-II also had a cadence of a few days, before taking into consideration bad weather, so would typically miss transients that evolve on timescales of $\sim\rm days$ \citep{Sullivan2011,Kessler2009,Miknaitis2007}. Although these surveys sample volumes much larger, the comparatively high cadence observations of \textit{Kepler}/\textit{K2} mean its legacy data can provide a unique wide-field transient survey. 

A worldwide push is being made to extend surveys into high cadences to explore a new time domain for transients. \citet{Andreoni2020} present one such effort, known as Deeper, Wider, Faster (DWF), which coordinates multi-messenger observations from 40+ telescopes for short $\sim1$~hour campaigns, with a cadence of 1.17~minutes. From this unique data set, an upper-bound on extragalactic fast transients was found to be $R_{eFT}< 1.625\rm ~ deg^{-2}\, d^{-1}$. Although \textit{Kepler} has a longer cadence of 30~minutes, a field is examined for $\sim$74~days. Similarly, the $TESS$ mission is an ongoing space-based mission, with 30-minute cadence and a wider field, albeit a lower sensitivity than \textit{Kepler}/\textit{K2} \citep{Ricker2014}.  \citet{Sharp2016} and \citet{Ridden-Harper2017} present a future high altitude balloon-based telescope system that aims to explore the time domain of days at ultraviolet wavelengths. 

\citet{LSSTScienceCollaboration2017} presents the case for high cadence observations, highlighting the unique short duration events expected to be seen at \textit{Kepler}-like cadences, such as GRB afterglows, other relativistic events, and supernovae shocks. As the time domain at the timescale of hours is relatively unexplored for transients, there is potential to discover exotic new transients and provide well-defined rates for current and upcoming high cadence transient surveys. Although \textit{Kepler}/\textit{K2} data can probe this parameter space, the classification of transients will be particularly challenging, since only observation in the \textit{Kepler} filter will be available for most candidates. 

Although challenging, examining every pixel in \textit{K2} data presents an opportunity to conduct a unique, high cadence survey over an appreciable volume. With this volume, \textit{Kepler} can be used to examine expected rates of exotic short duration events, such as binary neutron star mergers, similar to the analysis \cite{Scolnic2017} conducted on existing survey data. 

In this paper, we present the details of the ``\textit{K2}: Background Survey (\textit{K2}:BS)" along with examples of positive detections. A companion paper \citet{Ridden-Harper2019} presents the analysis of KSN-BS:C11a, the first transient discovered in \textit{K2}:BS. The analysis method is presented in \S~\ref{sec:method}, followed by the survey characteristics in \S~\ref{sec:surveycharacteristics} and example detections of known objects in \S~\ref{sec:examples}. In \S~\ref{Sec:Rates} we present the extragalactic volume surveyed and the expected detection rates for an assortment of transients in \textit{K2}.

\section{\textit{Kepler}/\textit{K2} data} \label{sec:data}
The \textit{Kepler}/\textit{K2} data provides high cadence photometry on a wide range of targets. In each of the 19 campaigns between 50--100 targets were observed in short cadence mode with frames every $\sim1$~minute, while between 10,000--20,000 targets were observed in long cadence mode, with frames every 30~minutes\footnote{https://keplerscience.arc.nasa.gov/data-products.html}.  In this paper, we will focus on data from the long cadence mode, however, this analysis technique is also applicable to short cadence data. Observation campaigns nominally lasted $\sim80$~days, with some campaigns shortened, due to technical difficulties.

\textit{Kepler} had a 116~deg$^2$ field-of-view (FOV), with a plate scale of 4\arcsec\  pixel$^{-1}$. Due to memory constraints, the entire \textit{Kepler} FOV could not be telemetered so instead, science targets were pre-selected each observing campaign and allocated a pixel mask that extended several pixels around the science target, both for background subtraction and to account for telescope drift. These pixel masks are known as target apertures in the downloaded data cube, the Target Pixel Files (TPFs), and all targets have unique identifiers, known as the EPIC number for \textit{K2} \citep{Huber2016} and KIC number for \textit{Kepler} \citep{Brown2011}. As the TPFs record both spatial and temporal information they are the data we analyse in this project. 

A key difference between \textit{K2} data and that of the original \textit{Kepler} mission is the poor telescope stability. Following the failure of 2 reaction wheels, leaving \textit{Kepler} with only 2 functional reaction wheels, the \textit{K2} mission was developed, using solar pressure on the solar array, in conjunction with the two remaining reaction wheels \citep{Howell2014}. This configuration was subject to drift up to $\sim1$ pixel (4\arcsec) every 6 hours, so the pointing was corrected by thruster firings every 6 hours, and are recorded as quality flags in the TPF. The drift is recorded in the TPF through two displacement parameters labelled POS\_CORR, which give the local image motion, calculated from fitting motion polynomials to the centroids of bright stars in each detector channel \citep{DAWGDataAnalysisWorkingGroup2012}. The POS\_CORR values give a valuable reference to telescope stability throughout campaigns, with the total image displacement given by $D^2 = \rm POS\_CORR1^2 + POS\_CORR2^2$.

All \textit{K2} fields are pointed along the ecliptic, enabling a range of galactic and extra-galactic fields to be observed. A consequence of pointing along the ecliptic is that a large number of asteroids cross through the \textit{K2} fields and contaminate data.

\subsection{\textit{Kepler/K2} magnitudes and zeropoints}
We convert Counts, $C$, observed by \textit{Kepler} to AB magnitudes, or $K_p$, according to
\begin{eqnarray}
K_p = \rm -2.5log(C) + z_p, \label{eq:mag}
\end{eqnarray}
where $z_p$ is the zeropoint. Since we analyse data across the entire \textit{Kepler/K2} field of view, we calculate the zeropoint independently for each of the 84 readout channels. This approach accounts for variability that may occur between the channels.

\begin{figure}
	\centering
	\includegraphics[width=\columnwidth]{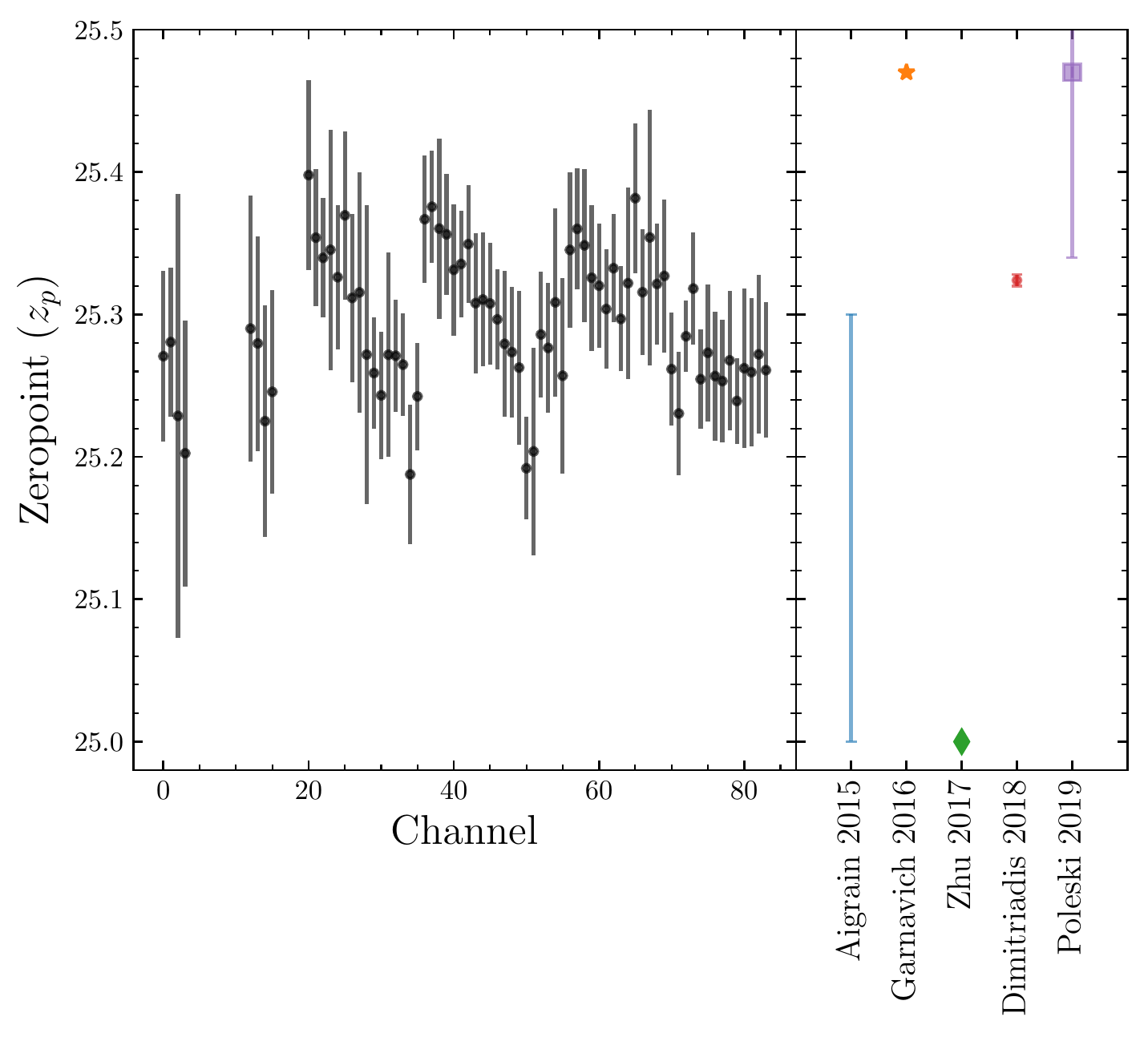}
	\caption{\textbf{Left:} Zeropoints we calculate for each \textit{Kepler/K2} channel, using stars from Campaigns C01, C06, C12, C16, and C17. \textbf{Right:} A selection of \textit{Kepler/K2} zeropoints taken from \citet{Aigrain2015}, \citet{Garnavich2016}, \citet{Zhu2017}, \citet{Dimitriadis2018}, and \citet{Poleski2019}. The complete list of zeropoints are presented in Table~\ref{tab:zeropoints}, and Table~\ref{tab:otherzp}.}
	\label{fig:zp}
\end{figure}

We calculate the $z_p$ for each channel by combining the $K_p$ magnitudes defined in the Ecliptic Plane Input Catalog \citep{Huber2016} with the de-trended PDC EVEREST light curves \citep{Luger2016} for all observed stars. We cut all stars with light curve variability $>1\%$, and then perform a 3$\sigma$ sigma clip to remove outliers. Finally, we calculate the zeropoint for each channel using Eq.~\ref{eq:mag} and average across campaigns C01, C06, C12, C14, C16, and C17. The zeropoints for each channel are shown in Fig.~\ref{fig:zp} with values listed in Table~\ref{tab:zeropoints}.

We find that the zeropoints vary up to $0.2$~mag across the channels. When compared to zeropoints derived by others, we find that our zeropoints are consistent, with the exception of \citet{Zhu2017}, which defines $z_p=25$. \citet{Ridden-Harper2020} builds on this work to establish \textit{Kepler/K2} zeropoints that are comprehensively calibrated using Pan-STARRS photometry.

\section{Methods} \label{sec:method}

The unique dataset from \textit{K2} presents a number of challenges for event detection. Spacecraft drift, throughout \textit{K2} data, presents the largest challenge, requiring special treatment. Conventional methods such as image subtraction proved to be ineffective with the \textit{K2} data. The motion of up to $\sim 4$\arcsec\ in \textit{K2} images resulted in poor subtractions with a median image. To counteract this we trialled subtracting images with similar displacements. While it achieved cleaner subtractions, it was not applicable to all images and introduced temporal biases in event detection. As these subtraction methods failed to be generally applicable, we developed new methods for event detection which still contain elements of conventional image subtraction.

We developed two methods to identify events, lasting from 1.5~hours to tens of days. One method can detect short events, with lifetimes $<10$~days, by assuming that the transient will be significantly brighter than background scatter; while the other method can detect long events with lifetimes $>10$~days, through heavy smoothing of the light curves and careful reference frame selection. The combination of these two methods successfully recovers short events, like asteroids and stellar flares, and long events such as supernovae.

The following section will describe the main data manipulation methods, followed by a detailed description of the detection methods. The analysis presented here is applicable to all \textit{Kepler}, \textit{K2}, and \textit{TESS} data.

\subsection{Science target mask} \label{sec:mask}

Each TPF contains science targets, other persistent sources and background pixels. To assist in event sorting, we identify each source and its extent. Identify the sources with a science target mask allows us to determine if events are associated with known objects, such as stars or galaxies.

We create the science target mask using a median frame with the following method. The median frame is generated by averaging a set of successive frames that have a total displacement less than 0.2~pixels from nominal telescope pointing. A sigma cut is performed on the median frame, where all pixels that are brighter than $\rm median+\sigma$ of all pixels in the median frame are added to the science target mask. The previous sigma cut is repeated with the newly defined science target mask applied to the median frame, all pixels identified in this sigma cut are added to the science target mask. 

To prevent transients from being included in the science target mask, we create two masks, one at the start and another at the end of the campaign. All science target mask pixels that appear in both the start and end masks are included in a final science target mask. Transients that can be identified by \textit{K2}:BS must be shorter than the \textit{K2} campaign duration ($\sim 80$~days), so this method ensures that detectable events aren't included in the science mask.

The large ($4\arcsec$) pixels of \textit{Kepler} make is possible that multiple sources blend together into a single mask. In an effort to separate these possible science targets, we separate the science target mask into multiple masks, through a watershed algorithm \citep{Barnes2014}. All of the science targets are checked against NASA/IPAC Extra-galactic Database (NED)\footnote{The NASA/IPAC Extra-galactic Database (NED) is operated by the Jet Propulsion Laboratory, California Institute of Technology, under contract with the National Aeronautics and Space Administration.} and the Simbad \citep{Wenger2000} database, to identify the object within each science target. 

\subsection{Telescope drift correction} \label{sec:drift}
During \textit{K2} the telescope drift introduced significant instrumental artefacts to the data. During campaigns, \textit{Kepler} drifts to a maximum of $\sim$1~pixel from nominal telescope pointing, over $\sim6$~hours. For bright targets, this effect is often negligible and can be offset with the motion correction tools present in Pyke \citep{pyke,pyke3}, Lightkurve \citep{lightkurve}, K2SFF \citep{Vanderburg2014}, and EVEREST \citep{Luger2016}. 

The techniques used with K2SFF and EVEREST were developed to primarily correct stellar light curves, that don't feature strong variability. As a result these techniques fail to correct the background motion, without removing the transient signal. The Lightkurve PRF photometry tool can correct for telescope motion, while preserving transient signal, however, PRF photometry currently requires tailoring priors, such as position and flux, for each object in the TPF, and has a long computation time. These issues make PRF photometry infeasible for this analysis. As no existing detrending methods are suitable for a transient search, we develop a method to fit and remove motion induced noise for each pixel.

\begin{figure}
    \includegraphics[width=\columnwidth]{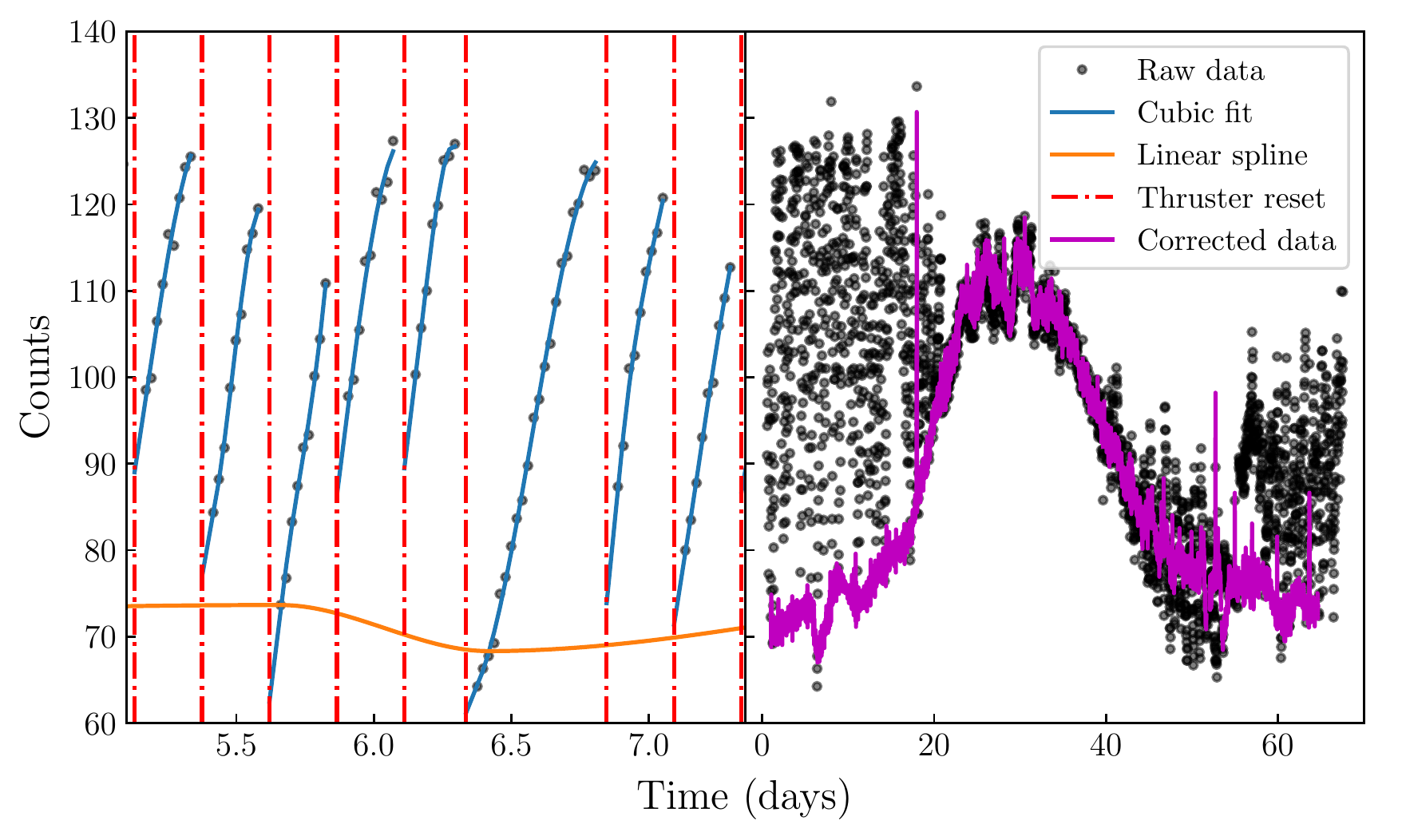}
    \caption{An example of the motion correction procedure on a pixel containing SN~2018ajj \citep{2018ATel11661....1S}. \textbf{Left:} The components of the fitting procedure are shown: the thruster resets (red dashed line) divide the light curve into segments; a linear spline (orange line) is fit to the most stable points in each light curve segment; a cubic polynomials (blue lines) are fit to the raw data (black points). \textbf{Right:} The full light curve is shown for SN~2018ajj, with raw data (black points), and the motion corrected data (magenta line). This process dramatically reduces noise induced by telescope motion.}
    \label{fig:MC}
\end{figure}

The telescope drift results in bright sources to periodically contaminating neighbouring pixels, creating discontinuous light curves. The impact that the telescope drift has on a light curve can be seen with the blue points in Fig.~\ref{fig:MC}. These motion induced signals increase the detection threshold and can masquerade as real transient signals, so it is critical that they are corrected. 

Our data reduction method relies on thruster resets, which are identified through the quality flag $1048576$. Utilising the thruster resets, we employ a data correction method that analyses each segment between thruster resets. The procedure acts on every pixel individually, altering the data with the following steps:
\begin{enumerate}

\item All images/frames that have $< 0.3$~pixel displacement in a campaign are identified.\footnote{This choice is made based on data quality and the number of observations available at high precision.} 

\item A 1D spline is fitted to the flux values of the image with the least displacement between thruster resets of the previously identified frames for each pixel.

\item The 1D splines are subtracted from each of the pixel light curves for the entire campaign, leaving residual flux that is largely the product of telescope motion.

\item The residual light curve is broken into segments defined by thruster resets.

\item To avoid real transients from being removed in the motion correction, we perform a $2\sigma$ sigma clip on the data.

\item For each residual light curve segment, we fit a cubic polynomial through linear regression with \texttt{scikit-learn} \citep{Varoquaux2015}.

\item The cubic polynomial fits are subtracted from the residual flux calculated in step iii.

\item Finally the the 1D spline is added back to the data, resulting in motion corrected data. 
\end{enumerate}

This procedure is applied to all pixels, where the correction is most prevalent in pixels near the science target, or field stars. A dramatic example of this correction can be seen in Fig.~\ref{fig:MC}, where through this motion correction we successfully reconstruct the light curve of SN~2018ajj \citep{2018ATel11661....1S}. No extrapolation is included in this method, so the data becomes truncated to the first and last instance where the pointing accuracy is $<0.3$ pixels.

We find this procedure is successful at removing almost all motion artefacts from pixel light curves. The only residual artefacts that have been encountered are ones that persist between multiple thruster resets. Although we do not remove these trends, the false positives they produce in the event detection are removed through vetting steps discussed in \S~\ref{sec:short_event_id}.

\subsection{Short event identification (< 10 days)} \label{sec:short_event_id}
The core aspect of \textit{K2}:BS is identifying real events from noise or other contaminants (e.g. asteroids). For each campaign every pixel is analysed independently to identify a potential event and an associated event time. A detection limit is calculated for each pixel as $Limit = \rm median + 3\sigma$ from all images. This cut-off value sets the magnitude limit for each pixel. Due to telescope drift, pixels close to science targets or stars have brighter limits than pixels with large separations. This highly variable level of counts, as seen in the magnitude limits shown in Fig.~\ref{fig:Maglim}, is the motivation for analysing pixels individually.

\begin{figure}
	\centering
	\includegraphics[width=\columnwidth]{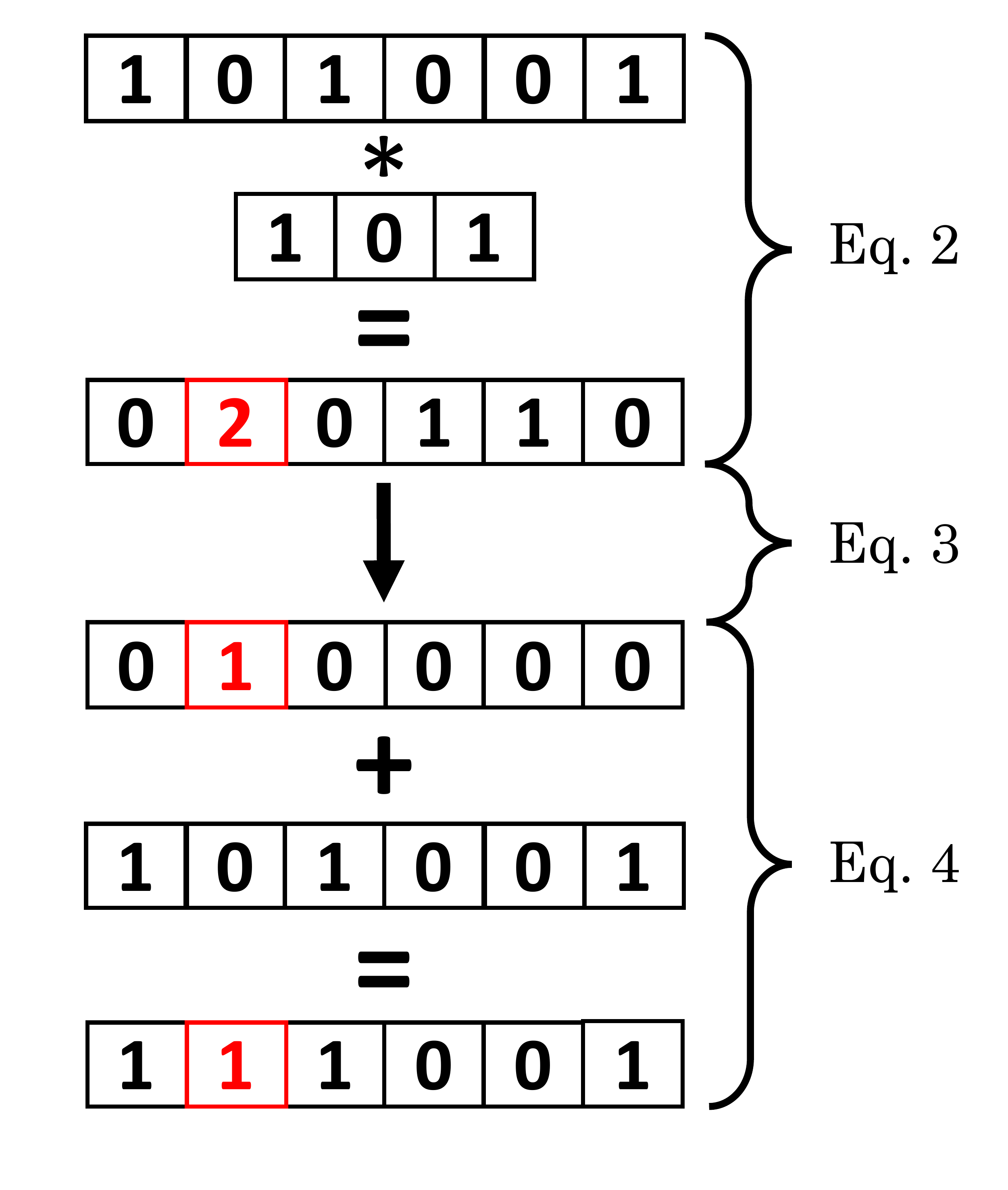}
	\caption{An example Boolean detection array being smoothed by a length 3 kernel. The smoothing prevents events from being truncated and rejected due to pixel count drops from instrument error.}
	\label{fig:smoothedbool}
\end{figure}

The process of identifying potential events utilises Boolean arrays and indexing. The \textit{K2} target pixel file flux is converted to a Boolean array, conditioned on pixels having counts greater than the aforementioned limit. To avoid anomalous detections based on spacecraft operation all pixels in frames that coincide with nonzero quality flags, indicating anomalous behaviour, are set to false. Residual telescope motion and noise can cause false breaks in the Boolean array, as the light curve may periodically drop below the detection limit. These breaks will produce false event durations and may lead to transients being processed incorrectly and lost. To prevent noise from truncating event duration the Boolean detection array is smoothed by an iterative process of convolving each pixel, through time, with a 1d kernel of zeros with ones at the start and end positions. The smoothing process iterates from a length 9 kernel, to a length 3 kernel, with each iteration acting on the product from the previous step. The convolution process is as follows,

\begin{eqnarray}
	C &=& D * k\\
	A(C) &=& 
	\begin{cases}
	1, & \text{if } C = 2\\
	0, &  \text{if } C < 2
	\end{cases}\\
	R &=& D + A(C),	
\end{eqnarray}

\noindent
where $D$ is the initial Boolean detection array, $k$ is the 1d smoothing kernel described above, $A(C)$ is the construction of a new Boolean array from, $C$, the convolved array, and $R$ is the final smoothed Boolean array used for detection. An example of this process with a length 3 kernel is shown in Fig.~\ref{fig:smoothedbool}.

We use the smoothed Boolean array to identify candidate events with durations longer than a chosen baseline. The event identification process analyses each pixel independently and operates as follows:
\begin{enumerate}
    \item We identify the indices of all False values in the Boolean array. \\
    \item The difference between neighbouring False indices is calculated. \\
    \item Differences that are greater than a predefined baseline are selected as candidate events. \\
    \item The start and end times of the candidate event are recorded along with the pixel position.
\end{enumerate}
we set the minimum length to be 3 frames, or 1.5~hours, this requirement avoids detecting spurious noise or cosmic rays. This method proves successful in identifying transients that evolve rapidly over 1.5~hours to $\sim10$~days.

Many of the potential events selected are false detections and contamination from known variable sources, that must be removed by subsequent checks. Sources of contamination can either be astrophysical (e.g., variable stars, asteroids), or instrumental (e.g., residual telescope motion). We impose a multi stage vetting procedure to limit false detections. 

First we match all coincidental events. For each candidate event we collapse the Boolean detection array along the time axis from the beginning to the end of the candidate events duration. We then construct a Boolean mask array and set the pixel position of the candidate event to be True. The mask is then iterative convolved with a $3\times 3$ True array and any True elements that appear in both collapsed Boolean array and the convolved mask are added to the mask, for the next iteration. This process terminates when no new pixel is added to the Boolean mask. Candidate events that are included in this new mask are merged into a single candidate.

Following the event matching procedure we vet candidates based on their the uniqueness. We smooth the light curve by with a Savitzky-Golay filter, setting the width to be twice the event duration. All peaks in the light curve are then identified using the \texttt{Scipy find\_peaks} algorithm, which identifies local maxima through neighbour comparisons, and we condition on the peaks. We assume a real event will be significant and unique, so candidate events are rejected if: \textbf{1)} the largest peak occurs outside the identified event time; \textbf{2)} all peaks within a time frame less than twice the event duration from the candidate event must be less than $80\%$ the brightness of the maximum peak. We find that in areas of prolonged poor pointing, a forest of similar peaks emerge, which condition 2 is instrumental in vetting. 

Although these conditions are successful in eliminating almost all false detections, they introduce complexity into the true limiting magnitude. A thorough analysis of the magnitude limits for each pixel will be presented in future work.

\subsection{Long event identification (> 10 days)} \label{sec:LongID}
This method closely follows that of the short detection method, however, greater care is taken in the construction of the pixel limits. As the \textit{K2} campaigns nominally last for $\sim80$~days, long transients, such as supernovae (SN) can strongly impact the light curve throughout the campaign, raising the detection limit to a point where the transient is no longer detectable. In order to construct an accurate limit we must examine trends in the light curve.

First, we smooth the light curve with a Savitzky-Golay filter with a window of 5~days. Although this process truncates the light curve by 10~days, it removes the signatures of short events (e.g., asteroids) from the light curve. We then identify the largest peak in the data using the \texttt{Scipy find\_peaks} algorithm \citep{Virtanen2019}, and break the light curve into 3 zones: the event zone, extending 10~days before to 25~days after the peak; and zones 1 and 2, before and after the event zone respectively. As some transients, such as SN can have long declines, we compare the light curve properties of zones 1 and 2 with the following; 
\begin{eqnarray}
 \mu_2 > \mu_1 + 3\sigma_1,
\end{eqnarray}
where $\mu_{1,2}$ are the light curve means of zones 1 and 2, and $\sigma_1$ is the standard deviation of zone 1. If this condition is met then only zone 1 is used to calculate the limit, otherwise, both zones are used to calculate the limit. As with the short detection method, this limit is determined for each pixel individually as $Limit = median + 3\sigma$. 

With limits for each pixel, candidate long events are selected through a similar process to short events. Without smoothing the Boolean array, we locate events by the positions of False values. In this case we require the event duration to last longer than 10~days. For these events, we assume that they are unique and have well behaved peaks, as such candidate events are vetted by the following conditions: \textbf{1)} the largest peak occurs inside the identified event time; \textbf{2)} the peak can be well fit by a $3^{rd}$ degree polynomial and poorly fit by a $1^{st}$ degree polynomial, requiring the coefficient of determination ($R^2$) to be $>0.95$ and $<0.5$ respectively.

\subsection{Variable stars}
Bright variable stars and their associated overflow into columns of pixels can lead to false detections. These false detections are mitigated by checking the pixels neighbouring the event pixel, if one of the neighbours contain $>100,000$ counts the event is considered false and discarded. This value is chosen as it encompasses the apparent variability in the saturation level of the pixels. If the bleeding of the target occurs at a lower count rate than the cut-off, then it may also be contained in a ``Probable" event category. This category is discussed further in \S~\ref{sec:sorting}.

\subsection{Asteroids}
\begin{figure}
    \includegraphics[width=\columnwidth]{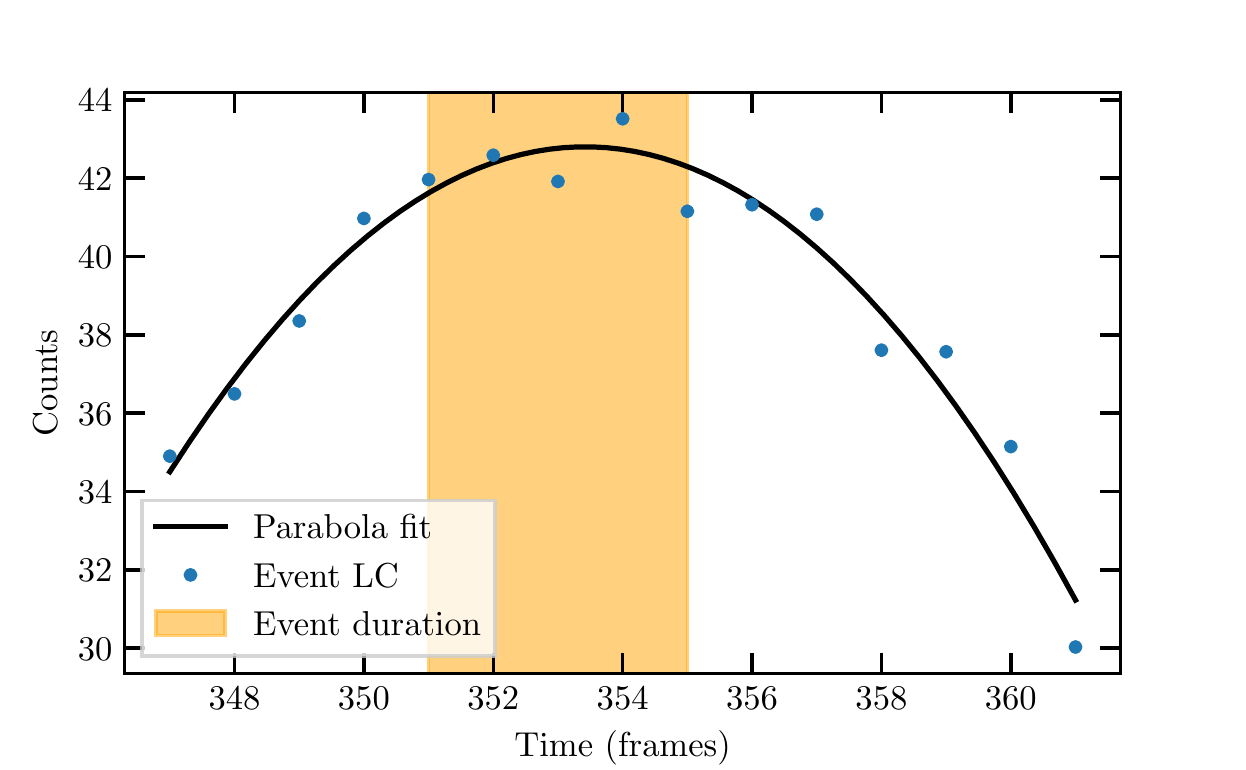}
    \caption{An example of a potential event identified as an asteroid in C01, EPIC 201735583. Since asteroids travel at a relatively constant speed through the aperture, the light curves they produce are largely parabolic. Thus parabolic light curves of short events are classified as asteroids.}
    \label{fig:Asteroid}
\end{figure}

Due to \textit{K2} observing along the ecliptic, there are many asteroids that pass through the data. As most asteroids remain in the TPF FOV for $<$~1~day, they are predominantly detected by the short event method. The asteroid detection method utilises the fact that asteroids tend to move with a near constant speed through the TPF, thus they produce a parabolic light curve, as seen in Fig.~\ref{fig:Asteroid}. \textit{K2}:BS identifies asteroids by fitting a parabola to frames $\pm 2$ from the brightest point in the potential event light curve. The generated parabola is then fitted and subtracted from the event light curve, if the residual is small then the potential event is categorised as an asteroid. All asteroids are recorded and can be examined independently. It is possible that real short transients may fall within the asteroid criteria, however, such events could be recovered at a later point following an analysis of asteroid-like events\footnote{Although uncommon, some asteroids will move through an epicycle within the TPF. These asteroids do not have uniform motion, and so can be missed by the asteroid filter. We are currently working on adaptive masks that are capable of tracking asteroids. These masks would extract all asteroid information and separate asteroids from transients.}.

\subsection{Event sorting} \label{sec:sorting}
For each object detected, we must search for a potential host or source of the event. To simplify visual vetting of candidate events, they are split into categories based on several event aspects such as duration, brightness, detection method, source type, and relation to masked objects. 

Identifying the host of a potential event can follow different pathways. For each detection, we query NED and Simbad to identify potential sources or host galaxies. If the event occurs within the mask of a science target, the host is defined to be the science target, and the candidate event is sorted into the ``In" category and subcategory based on the host type. Similarly, events that occur next to a science target mask are sorted into the ``Near" category and host type subcategory. Candidate events that occur in the background pixel, not associated with a science target, are identified through querying the coordinates corresponding to the brightest pixel, and sorted into a category corresponding to the likely host type. 

If a candidate event's coordinates do not correspond to an object in NED or Simbad database, it is assigned to the ``Unknown" category. This category often contains false detections form the \textit{K2} electronic noise sources, however it can also contain previously unseen events and some of the most exciting events promised by this analysis. An example of a new object found through an ``Unknown" classification can be found in \citet{Ridden-Harper2019}.

\subsection{Event ranking}
As a further diagnostic to assist in the vetting of candidate events is a series of quality rankings. Each candidate receives ranking for the brightness, duration, mask size, and source/host type. These rankings, as outlined below, provide a way of sorting candidate events into prioritised lists, simplifying the vetting process. 

\begin{itemize}
    \item The brightness ranking is taken as the significance of the event. The significance is found by calculating the peak flux of the candidate event and comparing it to the median and standard deviation of the entire light curve, excluding points from two days before the candidate event starts, to ten days after it ends. 
    \item The duration ranking is simply the calculated duration of the event in days. 
    \item The mask size ranking is calculated from the number of pixels included in the candidate event mask. This ranking is normalised such that three or more pixels in a mask produce a maximum rank of 1.
    \item The host ranking is based on classification given to the candidate event, from the NED object classification system. Objects that aren't of interest to this survey, such as stars, are ranked 0, while objects of interest, such as galaxies, quasars, etc., and unknown objects, are ranked 1. 
\end{itemize}

\subsection{Visual inspection}
\begin{figure*}
    \centering
    \includegraphics[width=\textwidth]{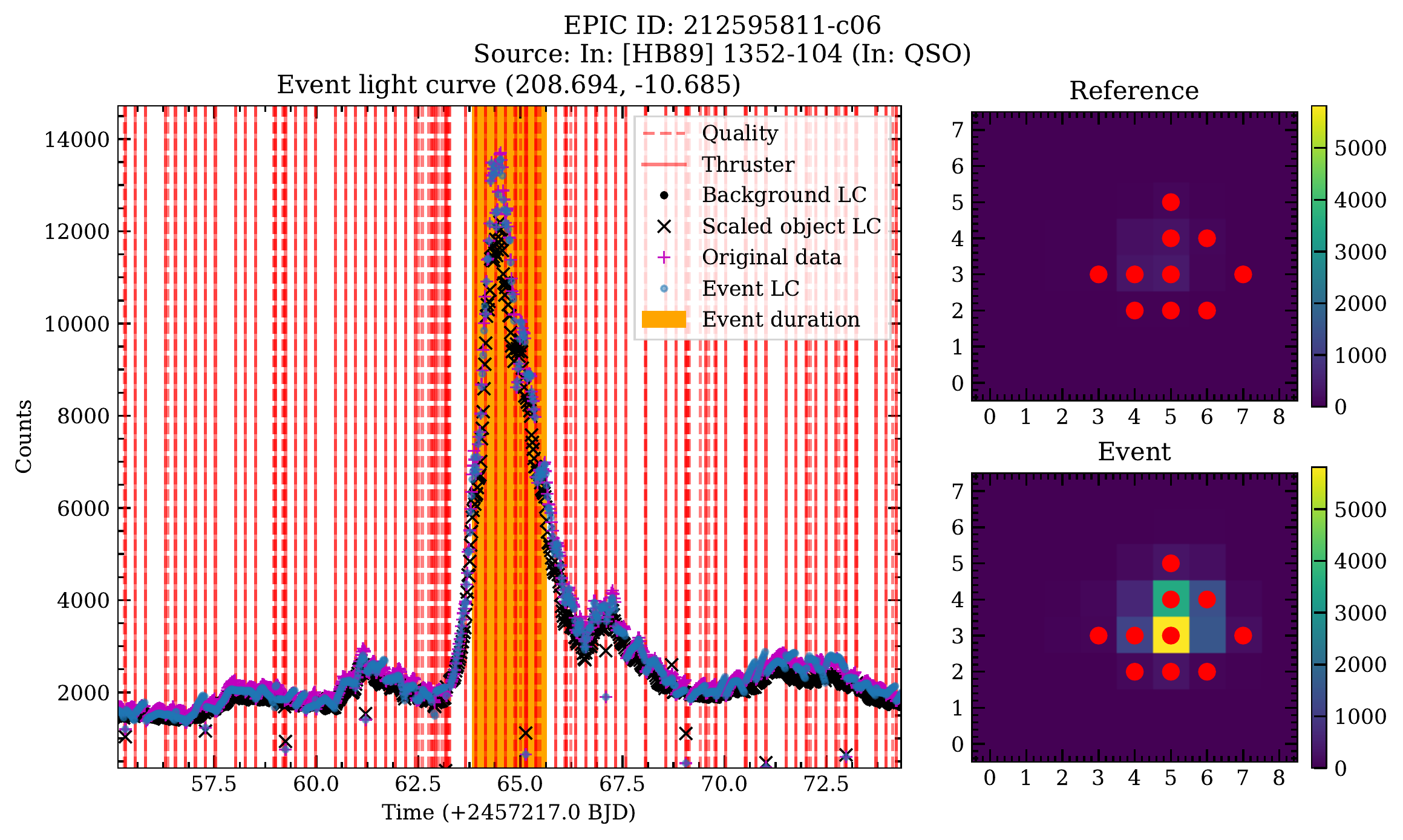}
    \caption{Example of a vetted, real event detected by \textit{K2}:BS. This event was detected in \textit{K2} object 212595811, which is a short 3 day outburst from the quasar [HB89] 1352-104. During the outburst the apparent magnitude increases from $17.2~K_p$ to $15.1~K_p$. The light curve on the left presents the event light curve along with diagnostic information such as when a thruster reset occurs and quality flags. The sub-figures on the right show the reference and peak brightness event frame, with the event mask overlaid in red points.}
    \label{fig:Visual}
\end{figure*}

If an event passes through the \textit{K2}:BS conditions it is finally checked through visual inspection. \textit{K2}:BS generates an event figure, video, and event positions. An example detection figure is shown in Fig.~\ref{fig:Visual} for a short outburst from quasar [HB89] 1352-104. The top of the figure contains diagnostic information on the object and position; left shows the event light curve, with diagnostic information, such as thruster firings and quality flags, the background and nearest science target light curves, and the identified event duration in orange; right shows the reference image at the top and the brightest frame from the event at the bottom.

\begin{figure}
    \centering
    \includegraphics[width=\columnwidth]{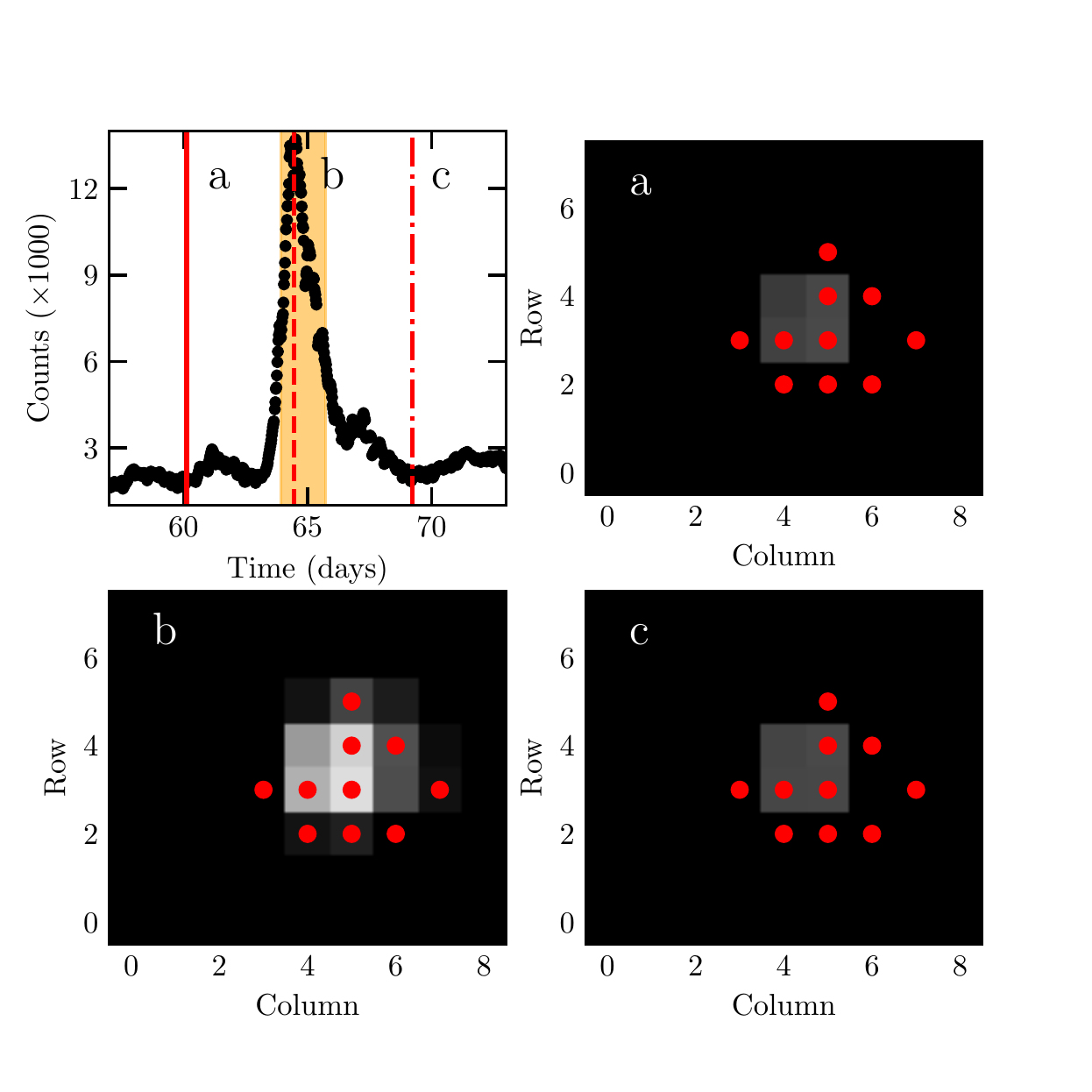}
    \caption{Example frames from the \textit{K2}:BS event video for a short outburst from quasar [HB89] 1352-104. A simplified light curve is shown in the top left, the detected event is highlighted in orange. Panels $a$, $b$, and $c$ are \textit{K2} images corresponding to the times indicated by red lines in the light curve.}
    \label{fig:movie}
\end{figure}

Selected frames from the corresponding event video for [HB89] 1352-104 are shown in Fig.~\ref{fig:movie}. Left is the event light curve with the vertical red line showing the time stamp of the \textit{K2} frame shown on the right.

These figures, alongside candidate event sorting provide enough information to accurately assess the validity of a candidate event.

\section{Survey Characteristics} \label{sec:surveycharacteristics}
As this survey is using existing data, survey limits and characteristics are set by the original data, as discussed in \S~\ref{sec:data}. Key survey parameters are the cadence, survey time, area, and depth. Operating with \textit{K2} long cadence data fixes the cadence to 30~minutes and survey time per field to $\leq80$~days. Events can only be detected if they are longer than the minimum event duration of 8~frames or 4~hours (set to prevent contamination from short scale noise or cosmic rays, and short enough to exhibit variation over a campaign, so that the long detection method can be successful.

Since two distinct detection methods are used that probe two different time domains, it is necessary to define the limits and recovery rates for both. Although the two methods target different time domains, there is an overlap for events with peaks or durations that last $\sim2$~days. This overlap prevents detection gaps in the time domain between the two.

As both detection methods scan every pixel, they feature the same survey area. For this analysis, we will focus on extra-galactic pointing fields, which have the best chance at detecting transients. As seen in Tab.~\ref{tab:area}, the background pixels cover a significant area. C01 features the highest number of background pixels, due to each science target being assigned larger pixel masks. Although C16 is an extra-galactic field, the area of background pixels is significantly lower than that of the other due to the field containing the Beehive cluster, and the Earth.

\begin{table}
 \caption{Number of background pixels, and the equivalent on sky area for \textit{K2} extragalactic pointing campaigns.}
 \label{tab:area}
 \begin{tabular}{lccc}
  \hline
  Campaign & Pixels & Area (deg$^2$) & Duration (days)\\
  \hline
  C01 & 3894581 & 4.8 & 83 \\
  C06 & 1957708 & 2.4 & 79 \\
  C12 & 2120713 & 2.6 & 80 \\
  C13 & 1645855 & 2.0 & 81 \\
  C14 & 2095376 & 2.6 & 81 \\
  C16 & 1408794 & 1.7 & 81 \\
  C17 & 2590027 & 3.1 & 69 \\
  \hline
 \end{tabular}
\end{table}

\subsection{Magnitude limits} \label{sec:limits}
\begin{figure} 
    \includegraphics[width=\columnwidth]{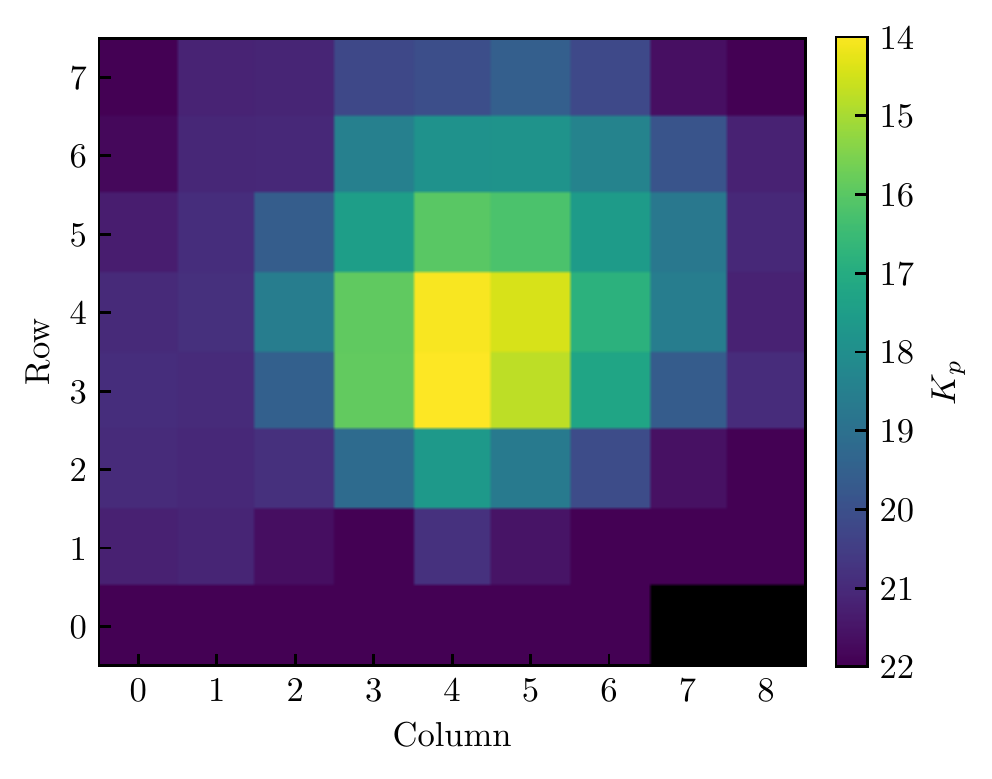}
    \caption{Magnitude limit for EPIC 212787678 from C17. Pixels near the original science target (centre frame) have a brighter limiting magnitude due to the science target's brightness, variability and residual telescope drift, with the pixels far from the target have a faint limiting magnitude. Each field has an associated magnitude limit from which the observed volume can be calculated.}
    \label{fig:Maglim}
\end{figure}

\textit{K2}:BS is limited to the area shown in Tab.~\ref{tab:area}, however, it has a highly variable magnitude limit. In this analysis the limiting magnitude is set by the detection limit that is imposed and discussed in \S~\ref{sec:short_event_id}~and~\ref{sec:LongID}. The interplay of the science target and telescope drift sets different magnitude limits, and therefore volumes, for each pixel. An example of the variable magnitude limit for a target pixel file can be seen in Fig.~\ref{fig:Maglim}. In general, pixels close to the science target have brighter limits, than those further away. This limit provides a simple way to calculate the expected volumetric rate of events. 

\begin{figure} 
    \includegraphics[width=\columnwidth]{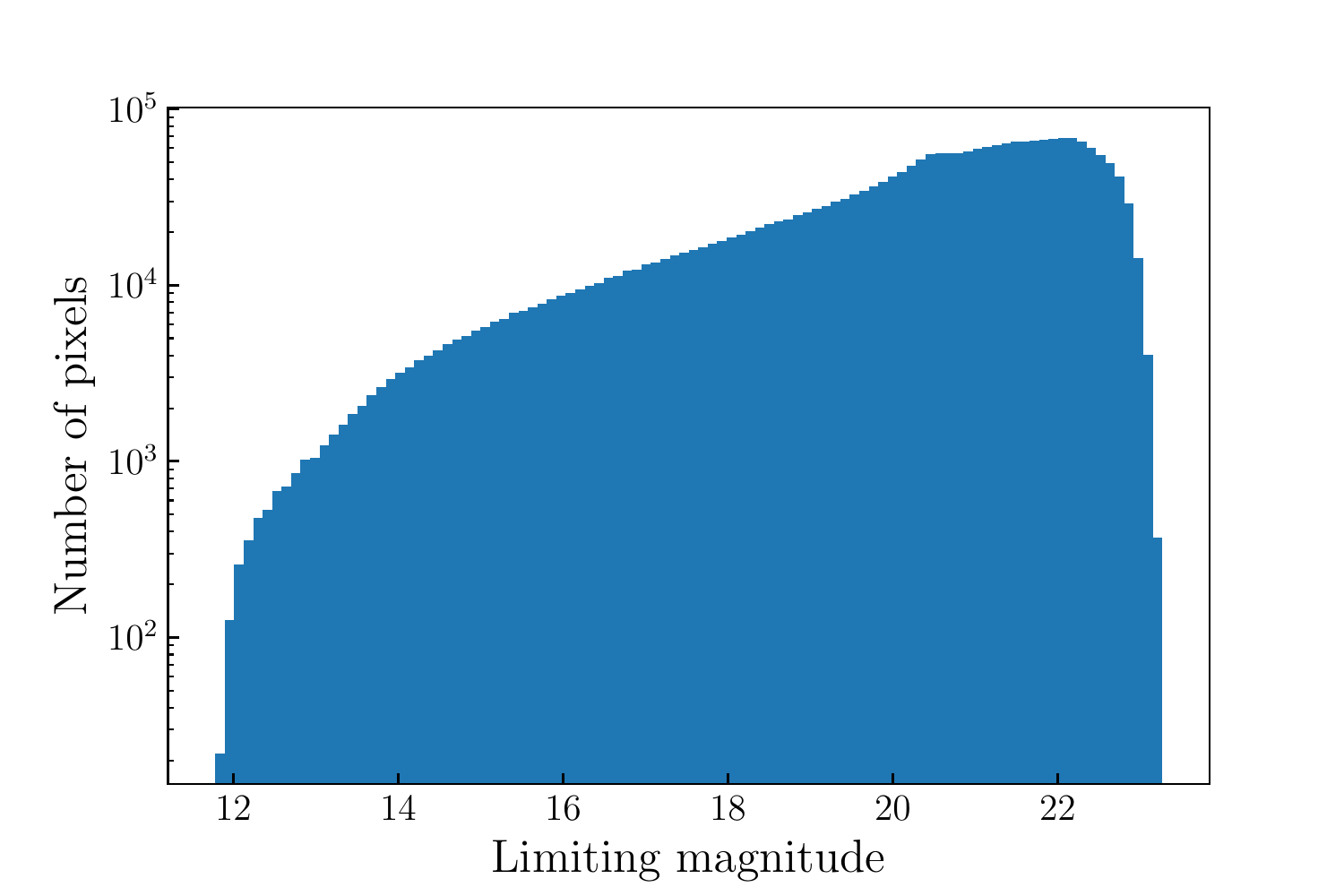}
    \caption{Distribution of pixel limiting magnitudes for C06. The high proportion ($\sim80$\%) of pixels with faint limits ($K_p\geq 18$) are ideal for transient searches.}
    \label{fig:limdist}
\end{figure}

Due to the mission design of \textit{K2}, the majority of downloaded pixels do not contain a target, and have faint limiting magnitudes. The distribution of pixel limiting magnitudes are shown for C06 in Fig.~\ref{fig:limdist}, where $\sim80$\% of pixels have limiting magnitudes $K_p\geq 18$, which are ideal for transient detection. Although some pixels report limiting magnitudes fainter than $K_p=22$, it is unlikely that such sensitivity can be achieved through this analysis. In all rate calculations we set the limiting magnitudes of pixels with $K_p>22$ to $22$.

Although the limit provided by \textit{K2}:BS is representative of the expected limiting magnitude, the candidate event vetting process will introduce complexities. As discussed in \S~\ref{sec:short_event_id}~and~\ref{sec:LongID} both the short and long detection methods undergo vetting to remove false detections. These checks have the potential to discard real transients if data quality is sub-optimal. We will present volumentric rates and a robust analysis of detection efficiency and contamination for a variety of key transients in a future work.

\section{Detected events} \label{sec:examples}
The pilot program of \textit{K2}:BS identified a number of known and undiscovered events. Initial runs have yielded promising results, by independently recovering key transients discovered in the \textit{Kepler} Extra-galactic Survey, and recovering a superoutburst of a dwarf WZ Sge nova which is presented in \citet{Ridden-Harper2019}. Here we will present example detections of known events, independently recovered by \textit{K2}:BS, that we use as method verification.

\begin{figure}
    \centering
    \includegraphics[width=\columnwidth]{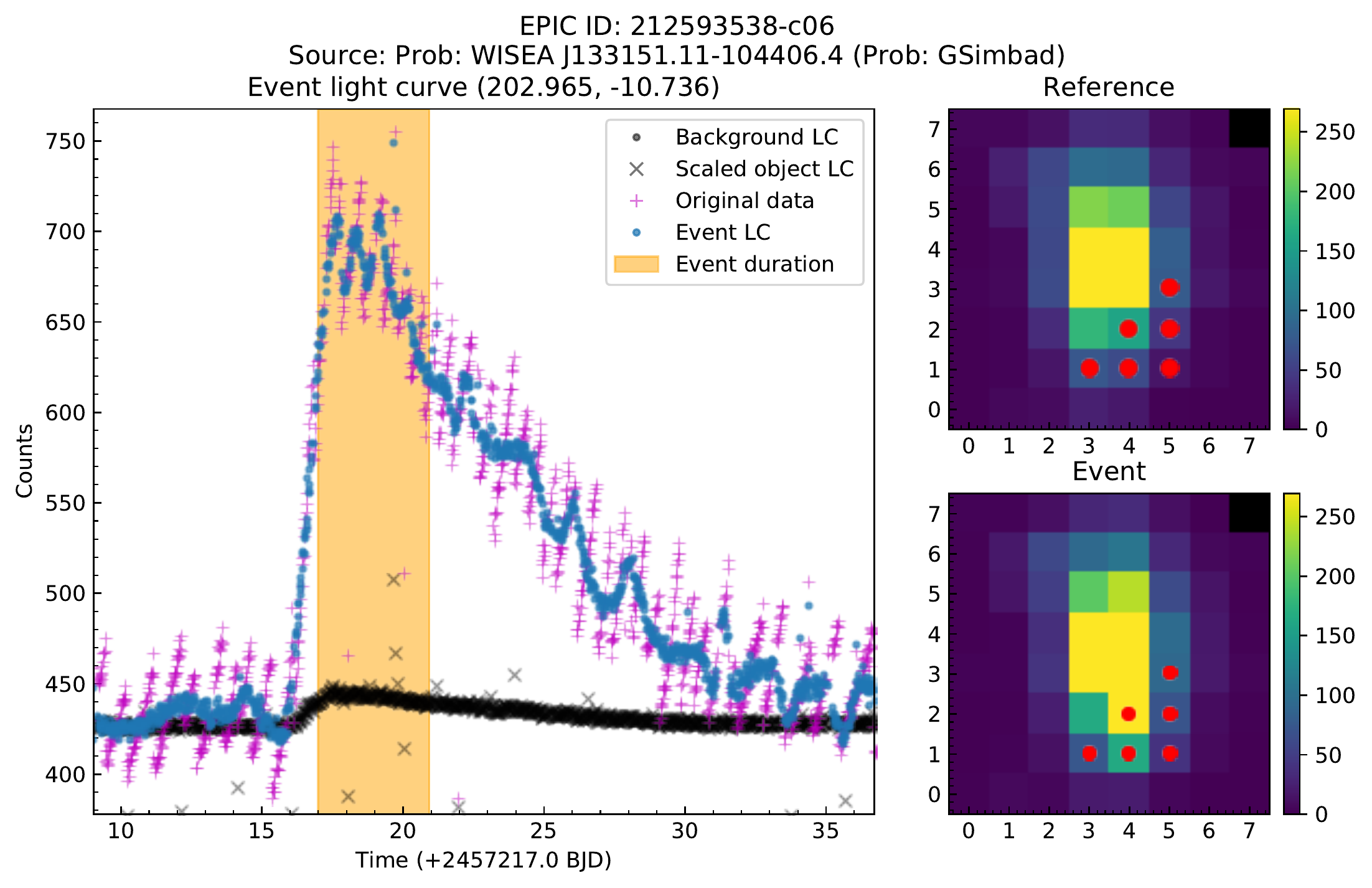}
    \caption{\textit{K2}:BS independent detection of KSN~2015K, presented in \citet{Rest2018}. This event was recovered through the ``short'' event detection method.}
    \label{fig:ksn2015k}
\end{figure}

Two examples of short transient events detected by \textit{K2}:BS are variability in quasar [HB89] 1352-104, and KSN~2015K. During C06, quasar [HB89] 1352-104 experienced an outburst that lasted $\sim3$~days, as seen in Fig.~\ref{fig:Visual}. Variability pre- and post- outburst are also visible. The short transient KSN~2015K \citep{Rest2018} was also recovered, as seen in Fig.~\ref{fig:ksn2015k}. The position of the event was determined to within to within 4\arcsec of the accepted value, which is the resolution limit of \textit{Kepler}. Both of these events are key examples of short events that may be detected by \textit{K2}:BS.

\begin{figure}
    \centering
    \includegraphics[width=\columnwidth]{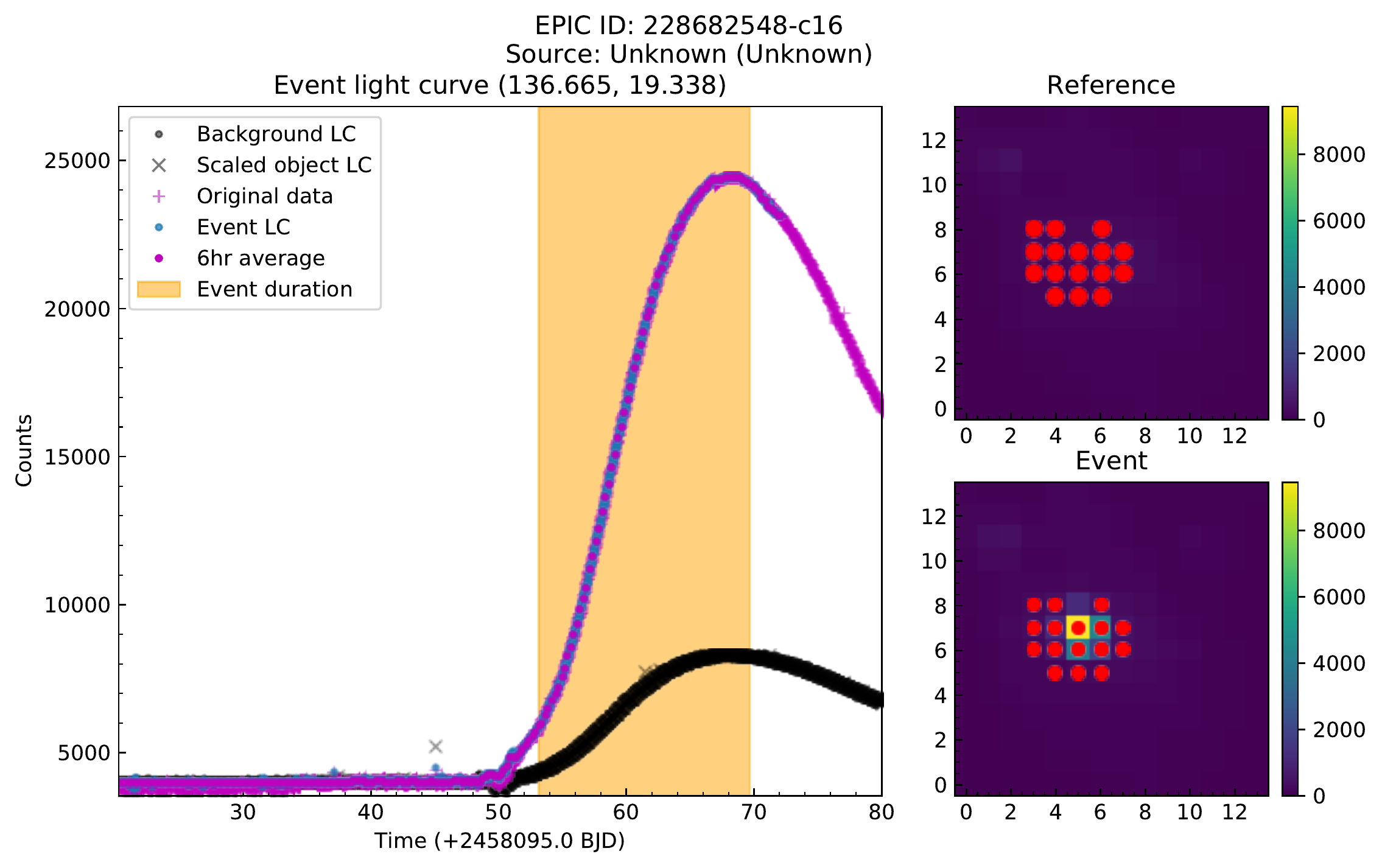}
    \caption{\textit{K2}:BS independent detection of SN~2018oh, presented in \citet{Dimitriadis2018,Shappee2018,Li2018}. This event was recovered with the ``long'' event detection method.}
    \label{fig:sn2018oh}
\end{figure}

\textit{K2}:BS is also capable of detecting type Ia supernovae, such as SN~2018oh, as seen in Fig.~\ref{fig:sn2018oh}. This event was detected through the long detection method. As well as detecting SN~2018oh, the excess light at rise is visible, as discussed in \citet{Dimitriadis2018,Shappee2018,Li2018}.

\section{Expected rates} \label{Sec:Rates}
The motivation for the \textit{K2}:BS is to detect transient events in \textit{K2} data, so its useful to determine the likely number of events the survey should detect. Since each \textit{K2} campaign is unique, the survey area and therefore event rates vary between campaigns. In the following sections we present the expected detection rate for SN Ia and CCSN, gamma ray burst (GRB) afterglows, kilonova, and the fast-evolving luminous transients (FELTs) detected by PAN-STARRS1 \citep{Drout2014}. FELTs, kilonova, and GRB afterglows evolve rapidly, so are detectable by the short detection method and we therefore use rates from this method, while  SN such as SN Ia and CCSN are detected through the long event detection method and we use respective rates.

There are few known types of transients that have lifetimes on the order of days. As \textit{K2}:BS will probe the $\sim1$~day time domain, the results of this survey will be used to define rates for day to sub-day duration transients and presented in future work.

\subsection{Volumetric rates}
For \textit{K2}:BS the volumetric rates are derived for each pixel and summed to reach the final rate. The motivation for this segmented approach is influenced by the large pixel size and, more importantly, the high variability in sensitivity between pixels. By knowing the magnitude limit per pixel we can calculate the volume, $V$, as follows,
\begin{eqnarray}
V = \frac{\theta}{3}\left(10^{\frac{1}{5}\left(m-M+5\right)} \right)^3 \nonumber
\end{eqnarray}
where $\theta$ is the solid angle of a pixel in steradians, $m$ is the magnitude limit of a pixel, and $M$ is the absolute magnitude of the transient. As these calculations are dependent on an absolute magnitude, the volumes are model dependent. 

For supernovae we use the \citet{Li2011a} type Ia supernova rate of $(0.301\pm0.062)\times 10^{-4}~\rm Mpc^{-3}\, yr^{-1}$ and an absolute magnitude of $-19.0$. For the kilonova rate we use $1540^{+3200}_{-1220}~\rm Gpc^{-3}\, yr^{-1}$ from \citet{Abbott2017d}, and take the absolute magnitude to be $-16$, based off GW170817 light curves presented in \citet{Villar2017}. The FELT rate is taken as $6400\pm2400~ \rm Gpc^{-3}\, yr^{-1}$ with an absolute magnitude range of $-16.5$ to $-20~K_p$ from \citet{Drout2014}. 

Since GRB light curves are highly model dependent, we will explore their rates in the following subsections.

\subsubsection{GRB rates}
For this analysis we use the \texttt{afterglowpy} package described in \citet{Ryan2019} for GRB afterglow light curves. In this exploratory case, we generate GRB afterglows using the \texttt{afterglowpy} top hat jet with the following set of fiducial GRB parameters; isotropic-equivalent energy, $E_0 = 10^{52}~\rm erg$; half-opening angle, $\theta_j = 3^\circ$ \citep{Guelbenzu2012}; circumburst density, $n=1.0~\rm cm^{-3}$; energy equipartition factors of the electrons and magnetic field, $\epsilon_e = 0.1$ and $\epsilon_B = 0.01$ respectively.

 The apparent magnitude of a GRB afterglow depends on both the viewing angle, $\theta_{obs}$, and the redshift, $z$. We calculate the apparent magnitude of the GRB afterglow independently varying $\theta_{obs}$ from $0-90^\circ$ in steps of $0.5^\circ$ and $z$ from $0.01-5$ in steps of $0.01$. By comparing the limiting magnitude of a pixel to the GRB magnitude array, we can determine the maximum $z$ a pixel could detect a GRB for a given viewing angle. 
 
 Following \citet{Guetta2005}, we approximate the GRB formation rate with the star formation rate from \citet{Rowan-Robinson1999};
 \begin{eqnarray}
 R_{GRB} = \rho_0 
 \begin{cases}
 10^{0.75z}, & z < 1 \\
 10^{0.75}, & z \geq 1
 \end{cases},
 \end{eqnarray}
 where $\rho_0 \approx 33~h^3_{65}~\rm Gpc^{-3}\,yr^{-1}$ \citep{Guetta2005}. From \citet{Japelj2011} the number of GRBs per year is given by
 \begin{eqnarray}
 N = \int_0^{z_{max}} \frac{R_{\rm GRB}}{1+z} \frac{dV}{dz}dz,
 \end{eqnarray}
 where the co-moving volume element,\quad $dV/dz$, is given by
 \begin{eqnarray}
 \frac{dV}{dz} = \frac{c}{H_0}\frac{4\pi\chi^2(z)}{E(\Omega_i,z)}, \quad \chi = \frac{c}{H_0}\int_0^{z'} \frac{dz'}{E(\Omega_i,z')},
 \end{eqnarray}
and $E(\Omega_m,z) = \sqrt{\Omega_m(1+z)^3+\Omega_\Lambda}$. For these calculations we adopt a standard $\Lambda\rm CDM$ cosmological model where $H_0=72~\rm km\,s^{-1}\,Mpc^{-1}$, $\Omega_m = 0.3$, and $\Omega_\Lambda = 0.7$.

With the fiducial GRB light curves and occurrence rate, we calculate the expected number of detections. The number of detections are calculated per pixel, accounting for all possible viewing angles. In these calculations we take ``On-Axis" GRBs to be $\theta_{obs}\leq \theta_{j}$ and ``Off-Axis" GRBs, or Orphan Afterglows, to be $\theta_{obs} > \theta_{j}$.

\subsection{Expected detection rates}
Tab.~\ref{tab:volrate} shows the expected number of events from various \textit{K2} extra-galactic campaigns. We find that it is unlikely that \textit{K2} serendipitously observed any kilonova or GRBs during the extragalactic pointing campaigns, however, it is likely that \textit{K2} serendipitously observed many SN~Ia and FELTs. Although the expected rates for SN~Ia and FELTs appear high, they are likely overestimates, due to intricacies in the \textit{K2} field selections. The \textit{Kepler} Extragalactic Survey (KEGS) survey will bias \textit{K2}:BS towards observing galaxies, as most large galaxies were selected, while the star targets may bias \textit{K2}:BS against galaxies. These subtle, but important, biasing factors will be incorporated into a follow-up paper where we present the comprehensive transient rates from \textit{K2}:BS, following completion of the survey. 

\begin{table*}
 \caption{Volumetric rates for the extra-galactic pointing \textit{K2} campaigns for SN~Ia, kilonovae, and FELTs. C01 has the largest rate, due to larger pixel masks around objects, and therefore more background pixels and the largest number of expected detections. Conversely C16, although an extra-galactic field, contains the Beehive cluster and Earth, which both severely reduce the sensitivity.}
 \label{tab:volrate}
 \begin{tabular}{lccccc}
  \hline
  Campaign & SN Ia & Kilonova (NS-NS) & FELT & \multicolumn{2}{c}{GRB}\\
             &                 & $\left(\times 10^{-3}\right)$ & $\left(-16.5~K_P\,\rightarrow\, -20.0~K_P\right)$ & On-axis & Off-axis\\
  \hline
  C01 & $12\pm2$ & $10^{+20}_{-8}$ & $0.08\pm0.02$ $\rightarrow$ $10\pm3$ & 0.14 & 0.03 \\
  C06 & $7\pm1$ & $6^{+11}_{-4}$ & $0.05\pm0.01$ $\rightarrow$ $6\pm1$ & 0.08 & 0.02 \\
C12 & $7\pm2$ & $6^{+12}_{-5}$ & $0.05\pm0.01$ $\rightarrow$ $6\pm2$ & 0.08 & 0.02 \\
C13 & $6\pm1$ & $5^{+9}_{-4}$ & $0.04\pm0.01$ $\rightarrow$ $5\pm1$ & 0.06 & 0.01 \\
C14 & $8\pm2$ & $6^{+13}_{-5}$ & $0.05\pm0.01$ $\rightarrow$ $6\pm2$ & 0.08 & 0.02 \\
C16 & $5\pm1$ & $4^{+8}_{-3}$ & $0.03\pm0.01$ $\rightarrow$ $4\pm1$ & 0.05 & 0.02 \\
C17 & $9\pm2$ & $7^{+15}_{-6}$ & $0.06\pm0.02$ $\rightarrow$ $8\pm2$ & 0.10 & 0.02 \\
  \hline
 \end{tabular}
\end{table*}

\section{Conclusions}
The \textit{Kepler} \textit{K2} mission has provided a wealth of information on transient events. Through a directed survey of over 40,000 galaxies KEGS found a multitude of transients, including a rare fast transient KSN 2015K \citep{Rest2018} and the potential detection of a SN~Ia shock interaction in SN~2018oh \citep{Dimitriadis2018,Shappee2018,Li2018}. The transients detected by KEGS are not the only transients detected by \textit{Kepler}, a number of events may have been detected serendipitously. The precedent on this was set by the discovery of a dwarf cataclysmic variable outburst in the original \textit{Kepler} mission \citep{Barclay2012}.

Through the \textit{K2}: Background Survey, we propose to analyse all pixels in \textit{K2} data for serendipitous transient detections. This survey will be blind, with final candidates vetted. The survey is not limited to specific types of transients, as all events that meet the detection threshold will be selected for final vetting. This unconstrained search, coupled with the rapid cadence of \textit{Kepler} will allow us to search for transients in the time domain of hours, which is poorly understood.

This survey will cover $\sim3\rm\,deg^2$ per campaign, with a total area of $\sim50\rm\,deg^2$ through all \textit{K2} fields. The total area, combined with the depth of up to 22 $K_p$ makes it likely that a number of unknown transients have been observed during \textit{K2}. This is made evident by the first reported detection of a WZ Sge dwarf nova superoutburst presented in \citet{Ridden-Harper2019}, discovered by \textit{K2}:BS. 

The analysis method presented here is not only applicable to \textit{K2} data. As \textit{Kepler} and \textit{TESS} share the same data structure as \textit{K2}, they too can be analysed through the background survey. The search strategies will also be more effective in \textit{Kepler} and \textit{TESS} data, due to the comparative high stability. Following the analysis of \textit{K2}, the background survey will run on these two additional data sets.

\section*{Acknowledgements}
We thank Geoffrey Ryan for his assistance in implementing \texttt{afterglowpy}. This research was supported by an Australian Government Research Training Program (RTP) Scholarship and utilises data collected by the K2 mission. Funding for the K2 mission is provided by the NASA Science Mission directorate. We also make significant use of the NASA/IPAC Extragalactic Database (NED), which is operated by the Jet Propulsion Laboratory, California Institute of Technology, under contract with the National Aeronautics and Space Administration.\\






\bibliographystyle{mnras}
\bibliography{AllPhD.bib}


\appendix
\section{\textit{Kepler/K2} zeropoints}
\begin{table*}
 \caption{Zeropoints for all \textit{Kepler/K2} detector channels. Channels with no values belong to the defunct modules 3, 4 and 7. The zerpoints are derived with data from Campaigns C01, C06, C12, C14, C16, and C17. }
 \label{tab:zeropoints}
 \begin{tabular}{lc|lc}
  \hline
  Channel (0-42) & Zeropoint (0-42) & Channel (43-84) & Zeropoint (43-84) \\
  \hline
  1 & $25.27\pm0.06$ & 43 & $25.35\pm0.04$ \\
2 & $25.28\pm0.05$ & 44 & $25.31\pm0.05$ \\
3 & $25.23\pm0.16$ & 45 & $25.31\pm0.05$ \\
4 & $25.2\pm0.09$ & 46 & $25.31\pm0.04$ \\
5 & $-$ & 47 & $25.3\pm0.04$ \\
6 & $-$ & 48 & $25.28\pm0.05$ \\
7 & $-$ & 49 & $25.27\pm0.05$ \\
8 & $-$ & 50 & $25.26\pm0.05$ \\
9 & $-$ & 51 & $25.19\pm0.04$ \\
10 & $-$ & 52 & $25.2\pm0.07$ \\
11 & $-$ & 53 & $25.29\pm0.04$ \\
12 & $-$ & 54 & $25.28\pm0.05$ \\
13 & $25.29\pm0.09$ & 55 & $25.31\pm0.07$ \\
14 & $25.28\pm0.08$ & 56 & $25.26\pm0.07$ \\
15 & $25.23\pm0.08$ & 57 & $25.35\pm0.05$ \\
16 & $25.25\pm0.07$ & 58 & $25.36\pm0.04$ \\
17 & $-$ & 59 & $25.35\pm0.05$ \\
18 & $-$ & 60 & $25.33\pm0.05$ \\
19 & $-$ & 61 & $25.32\pm0.04$ \\
20 & $-$ & 62 & $25.3\pm0.04$ \\
21 & $25.4\pm0.07$ & 63 & $25.33\pm0.04$ \\
22 & $25.35\pm0.05$ & 64 & $25.3\pm0.04$ \\
23 & $25.34\pm0.04$ & 65 & $25.32\pm0.07$ \\
24 & $25.35\pm0.08$ & 66 & $25.38\pm0.05$ \\
25 & $25.33\pm0.05$ & 67 & $25.32\pm0.04$ \\
26 & $25.37\pm0.06$ & 68 & $25.35\pm0.09$ \\
27 & $25.31\pm0.06$ & 69 & $25.32\pm0.04$ \\
28 & $25.32\pm0.08$ & 70 & $25.33\pm0.05$ \\
29 & $25.27\pm0.1$ & 71 & $25.26\pm0.04$ \\
30 & $25.26\pm0.04$ & 72 & $25.23\pm0.04$ \\
31 & $25.24\pm0.04$ & 73 & $25.28\pm0.03$ \\
32 & $25.27\pm0.07$ & 74 & $25.32\pm0.04$ \\
33 & $25.27\pm0.04$ & 75 & $25.25\pm0.03$ \\
34 & $25.26\pm0.04$ & 76 & $25.27\pm0.05$ \\
35 & $25.19\pm0.05$ & 77 & $25.26\pm0.05$ \\
36 & $25.24\pm0.04$ & 78 & $25.25\pm0.04$ \\
37 & $25.37\pm0.04$ & 79 & $25.27\pm0.05$ \\
38 & $25.38\pm0.04$ & 80 & $25.24\pm0.03$ \\
39 & $25.36\pm0.06$ & 81 & $25.26\pm0.06$ \\
40 & $25.36\pm0.04$ & 82 & $25.26\pm0.05$ \\
41 & $25.33\pm0.05$ & 83 & $25.27\pm0.06$ \\
42 & $25.34\pm0.04$ & 84 & $25.26\pm0.05$ \\

  \hline
 \end{tabular}
\end{table*}

\begin{table}
    \centering
    \caption{\textit{Kepler/K2} zeropoints from a selection of previous works.}
    \begin{tabular}{c|c}
    \hline
    Publication & Zeropoint \\\hline
    \citet{Aigrain2015}& $25-25.3$ \\
    \citet{Garnavich2016} & $25.47$ \\
    \citet{Zhu2017} & $25$\\
    \citet{Dimitriadis2018} & $25.324\pm0.004$\\
    \citet{Poleski2019} & $25.47\pm 0.13$ \\ \hline
    \end{tabular}
    
    \label{tab:otherzp}
\end{table}

\bsp    
\label{lastpage}
\end{document}